\documentclass[aps,12pt,showkeys,showpacs]{revtex4}
\usepackage{epsfig}

\newcommand{\cp}{\chi^{(+)}}
\newcommand{\cm}{\chi^{(-)*}}

\newcommand{\vv}{V_{bc}({\bf r}_1)}
\newcommand{\ri}{{\bf r}_i}
\newcommand{\ro}{{\bf r}_1}
\newcommand{\ak}{{\bf k}_a}
\newcommand{\bq}{{\bf k}_b}

\newcommand{\rc}{{\bf r}_c}
\newcommand{\cq}{{\bf k}_c}

\begin{document}


\title{Beam energy dependence of Coulomb-nuclear interference in the
breakup of $^{11}$Be}
%
\author{R. Chatterjee}
\email{rajdeep.chatterjee@pd.infn.it}
\affiliation{
Dipartimento di Fisica and INFN, Universit\`a di Padova, via F. Marzolo 8, I-35131, Padova, Italy}

\begin{abstract}
Within the
post-form distorted wave Born approximation 
wherein pure Coulomb, pure nuclear and
their interference terms are treated consistently in a single setup we study 
the beam energy dependence of the Coulomb-nuclear interference terms in the breakup of 
$^{11}$Be on a medium mass $^{44}$Ti target.  
Our results suggest that the Coulomb-nuclear interference terms are dependent on the incident beam energy and
can be as big as that of the individual Coulomb or nuclear terms depending on the angle and energy of the
breakup fragments.
We also calculate the relative energy spectra and one-neutron removal cross sections
in the breakup of $^{11}$Be on a heavy $^{208}$Pb target at 69 MeV/nucleon
for two different angular ranges of the projectile center of the mass scattering angle and compare them
with recently available experimental data.
\end{abstract}
\maketitle

\section{Introduction}
Nuclei away from the valley of stability have opened a new paradigm in nuclear
physics. They are often extremely unstable (especially those closer to the driplines) 
and have structure and properties which are quite often 
different from stable isotopes.
Many of them exhibit a halo structure
in their ground states in which loosely bound valence nucleon(s)
has (have) a large spatial extension with respect to the respective
core \cite{han01,tani,rii94,bert93}. 

There have been several attempts, within a fully microscopic approach, to understand the stability of these weakly bound systems. A few of them, like
different continuum shell model approaches \cite{csm3,volya}, including the shell model embedded in the continuum \cite{phyrep,cha06}, are formulated in the Hilbert space, i.e., they are based on the completeness of a single particle basis consisting of bound orbits and a real continuum.  A different approach to the treatment of particle continuum is proposed in the Gamow shell model \cite{Mic02}, which is the multi-configurational shell model with  a single particle basis given by the Berggren ensemble \cite{Ber68} consisting of Gamow (or resonant) states and the complex non-resonant continuum of scattering states. A microscopic cluster study of neutron rich carbon isotopes has also been
performed with the Generator Coordinate method in Ref. \cite{des_car}. However, a lot more needs to be done before one gets a more complete theoretical understanding of the underlying processes.

Another, widely used, method in unraveling the
structure and properties of halo nuclei, is to use breakup reactions featuring the exotic projectile. 
For reviews of this ever burgeoning field from an experimental and theoretical perspective one is 
referred to Ref. \cite{jon} and Ref. \cite{bau_p}, respectively.
It is now abundantly clear
that pure Coulomb \cite{nak94,nak99,pb1,pb2,cha00,cha01,pbc} 
or pure nuclear \cite{bhe,bon00,yab92}
breakup calculations may not be fully sufficient to describe all the 
details of the
halo breakup data which have been increasing rapidly both in quality and 
quantity \cite{han95,dav98,gil01,ritu02,udp,palit,fuku}. In majority of them both 
Coulomb and nuclear breakup effects as well as their interference terms are
likely to be significant and the neglect of the latter terms may not
be justified \cite{ann90,ann94,pb93,mad01}. 
The importance of nuclear effects even in the breakup of $^{8}$B in collisions with heavy
ions has been discussed in Ref. \cite{andrea}.
Thus, an important
requirement in interpreting the data obtained from the experiments 
done already or are planned to be done in future is to have
a theory which can
take care of the Coulomb and nuclear breakup effects as well as 
their interference terms on an equal footing. 

For breakup reactions of light stable nuclei, such a theory has been
developed \cite{bau84} within the framework of post-form distorted
wave Born approximation (DWBA), which successfully describes the 
corresponding data at low beam energies. However, since it uses the 
simplifying approximation of a zero-range interaction \cite{sat64}
between constituents of the projectile, it is inapplicable to cases
where the internal orbital angular momentum of the projectile is 
different from zero.  

Recently, we have presented a theory \cite{cha02,cha03} to 
describe the breakup 
reactions of one-nucleon halo nuclei within the post-form DWBA (PFDWBA)
framework, that  
consistently includes both Coulomb and nuclear interactions between the
projectile fragments and the targets to all orders, but treats the 
fragment-fragment interaction in first order. The Coulomb and nuclear
breakups along with their interference term are treated within a
single setup in this theory. The breakup contributions from the
entire continuum corresponding to all the multipoles 
and the relative orbital angular momenta between the valence
nucleon and the core fragment are included in this theory where
finite range effects are treated by a local momentum approximation
(LMA) \cite{shy85,bra74}. Full ground state wave function of the
projectile, of any angular momentum structure, enters as an
input to this theory.

The Coulomb-nuclear interference (CNI) terms have also
been calculated using the prior-form DWBA \cite{shy99} and within 
models \cite{typ01,bon02} where the time evolution of the 
projectile in coordinate space is described by solving the 
time dependent Schr\"{o}dinger equation, treating
the projectile-target (both Coulomb and nuclear) interaction as 
a time dependent external perturbation. 
Recently, the dynamical eikonal method, which unifies the semiclassical time dependent
and eikonal method and also takes into account interference effects has been used to calculate
realistic differential cross sections \cite{bay1,bay2}.
Coulomb and nuclear
processes have also been treated consistently on the same footing, in the continuum discretized
coupled-channels method \cite{cd1,cd2,cd3} and also within an eikonal-like framework,
in Refs. \cite{bon03,bon06}.
Nevertheless, to the best of our knowledge, the question of beam energy dependence of CNI has not been studied
within breakup models before. Thus there is a need to address this question and investigate this new physics
aspect within existing breakup theories itself, in view of several sophisticated experiments planned in the future.

In this paper, we investigate the beam energy dependence of the Coulomb-nuclear interference terms in the breakup of 
$^{11}$Be on a medium mass $^{44}$Ti target  
and also calculate the relative energy spectra and one-neutron removal cross sections
in the breakup of $^{11}$Be on a heavy $^{208}$Pb target at 69 MeV/nucleon
for two different angular ranges of the projectile center of the mass (c.m.) scattering angle.
 Our formalism is presented 
in section II. In section III, we present and discuss the results of our 
calculations for the breakup of $^{11}$Be on $^{44}$Ti and $^{208}$Pb targets.
 Summary and conclusions of our work  are presented
in section IV. 

\section{Formalism}

We consider the elastic breakup reaction, $a + t \to b + c + t$, in which the
projectile $a$ ($a = b +c$) breaks up into fragments $b$ and $c$ (both of
which can be charged) in the Coulomb and nuclear fields of a target $t$.
The triple differential cross section for this reaction is given by
\begin{eqnarray}
{{d^3\sigma}\over{dE_bd\Omega_bd\Omega_c}} & = &
{2\pi\over{\hbar v_a}}\rho(E_b,\Omega_b,\Omega_c)
\sum_{\ell m}|\beta_{\ell m}|^2, \label{eq1_5}
\end{eqnarray}
where $v_a$ is the relative velocity of the projectile with 
respect to the target, $\ell$ is the orbital angular momentum for the 
relative motion of $b$ and $c$ in the ground state of $a$, and 
$\rho(E_b,\Omega_b,\Omega_c)$ is the appropriate phase space factor (see,
e.g., Ref. \cite{cha00}).
The reduced transition amplitude, in Eq. (\ref{eq1_5}), $\beta_{\ell m}$ is
defined as 
\begin{eqnarray}
\hat{\ell}\beta_{\ell m}(\bq,\cq;\ak) & = & 
\int d\ro d\ri\cm_b(\bq,{\bf r})\cm_c(\cq,\rc) \vv \nonumber \\
& & \times u_\ell (r_1) Y^\ell_{m} ({\hat r}_1)\cp_a(\ak,\ri), \label{eq2_5}
\end{eqnarray}
with ${\hat \ell} \equiv \sqrt{2\ell + 1}$.
In Eq. (\ref{eq2_5}), functions $\chi_i$ represent 
the distorted waves for the relative motions of various particles 
in their respective channels with appropriate 
boundary conditions. 
The superscripts $(+)$  and $(-)$ represents 
outgoing and ingoing wave boundary conditions, respectively.
Arguments of these functions contain the
corresponding Jacobi momenta and coordinates.  $\vv$ represents the
interaction between $b$ and $c$, and $u_\ell (r_1)$ is the radial part of
the corresponding wave function in the ground state of $a$. 
The position vectors (Jacobi coordinates) satisfy the relations 
(see also Fig. 1 of Ref. \cite{cha00}): 
\begin{eqnarray}
{\bf r} &=& \ri - \alpha\ro,~~ \alpha = {m_c\over {m_c+m_b}},   \\
\rc &=& \gamma\ro +\delta\ri, ~~ \delta = {m_t\over {m_b+m_t}}, 
~~  \gamma = (1 - \alpha\delta),    
\end{eqnarray}
where $m_i$ ($i=a,b,c,t$) are the masses of various particles. In, what follows, we shall
recollect only those formulae which are essential for our discussion. More details of the theory,
especially those regarding the evaluation of $\beta_{\ell m}$, can be found in Ref. \cite{cha03}. 

The reduced amplitude, $\beta_{\ell m}$ [Eq. (\ref{eq2_5})], involves 
a six-dimensional integral which makes its evaluation quite complicated.
The problem gets further aggravated due to the fact that the integrand
involves the product of three scattering waves that exhibit an oscillatory
behavior asymptotically. In order  
to facilitate an easier computation of Eq.~(\ref{eq2_5}), 
we perform a Taylor series expansion of the distorted waves
of particles $b$ and $c$ about ${\bf r}_i$ and write 
\begin{eqnarray}
\chi^{(-)}_b(\bq,{\bf r}) & = & e^{-i\alpha{\bf K}_b.\ro}
                           \chi^{(-)}_b(\bq,\ri), \label{eq3_5} \\
\chi^{(-)}_c(\cq,{\bf r}_c) & = & e^{i\gamma{\bf K}_c.\ro}
                           \chi^{(-)}_c(\cq,\delta\ri).  \label{eq4_5}
\end{eqnarray}
Employing the LMA \cite{shy85,bra74}, the magnitudes
of momenta ${\bf K}_j$ are taken as 
\begin{eqnarray}
K_j(R)  = \sqrt {(2m_j/ \hbar^2)[E_j- V_j(R)]}, \label{ea6}
\end{eqnarray}
 where $m_j$ ($j=b,c$) is the reduced mass of the $j-t$ system,
$E_j$ is the energy of particle $j$ relative to the target in the
center of mass (c.m.) system, and $V_j(R)$ is the potential between $j$ and 
$t$ at a distance $R$. 
Finally, one obtains
\begin{eqnarray}
{\hat \ell} \beta_{\ell m} &=&
{(4\pi)^{3} \over {k_a k_b k_c\delta }} i^{-\ell} Y^\ell_{m_\ell}({\hat {\bf Q}})
Z_\ell (Q) \sum_{L_aL_bL_c} (i)^{L_a-L_b-L_c} {\hat L}_b{\hat L}_c
\nonumber \\
& \times & {\cal Y}^{L_b}_{L_c}({\hat k}_b,{\hat k}_c)
 \langle L_b 0 L_c 0| L_a 0 \rangle 
 {\cal R}_{L_b,L_c,L_a}(k_a,k_b,k_a), \label{eq17_5}
\end{eqnarray}
where
\begin{eqnarray}
{\cal Y}^{L_b}_{L_c}({\hat k}_b,{\hat k}_c) & = &
\sum_M (-)^M\langle L_b M L_c -M|L_a 0 \rangle
Y^{L_b}_{M}({\hat {k}}_b)Y^{L_c*}_{M}({\hat {k}}_c), \\  
Z_\ell (Q) & = & \int_0^{\infty} r_1^2 dr_1 j_{\ell}(Qr_1) u_\ell(r_1)V_{bc}(r_1),
 \label{eqst}\\ 
{\cal R}_{L_b,L_c,L_a} & = & \int_0^{\infty}
{{dr_i}\over {r_i}} f_{L_a}(k_a,r_i)
f_{L_b}(k_b,r_i) f_{L_c}(k_c,\delta r_i), \label{eq18_5} 
\end{eqnarray}
and ${\bf Q} = \gamma {\bf K}_c - \alpha {\bf K}_b$. In Eq.~(\ref{eq18_5}), $f_{L_i}$ $(i=a,b,c)$ are the radial part
of the partial wave $(L_i)$ expansion of distorted wave $\chi^{(\pm)}_i$, and is calculated by solving the Schr\"odinger
equation with proper optical potentials which includes both Coulomb and nuclear terms.
 Generally, the maximum value of the partial
waves $L_a,L_b,L_c$ must be very large in order to ensure the convergence
of the partial wave summations in Eq.~(\ref{eq17_5}). However, for 
the case of
one-neutron halo nuclei, one can make use of the following method 
to include summations over infinite number of partial
waves. We write $\beta_{\ell m}$ as
\begin{eqnarray}
\beta_{\ell m} & = & \sum_{L_i = 0}^{L_{i}^{max}} {\hat \beta}_{\ell m} (L_i) 
             + \sum_{L_i = L_{i}^{max}}^{\infty} 
                    {\hat \beta}_{\ell m}(L_i), \label{eq19_5}   
\end{eqnarray}
where ${\hat \beta}$ is defined in the same way as Eq.~(\ref{eq17_5}) 
except for
the summation sign and $L_i$ corresponds to $L_a$, $L_b$, and $L_c$. If
the value of $L_i^{max}$ is chosen to be appropriately large, the
contribution of the nuclear field to the second term of Eq.~(\ref{eq19_5}) 
can be
neglected and we can write
\begin{eqnarray}
\sum_{L_i = L_{i}^{max}}^{\infty}{\hat \beta}_{\ell m}(L_i)  \approx    
            \sum_{L_i = 0}^{\infty}{\hat \beta}_{\ell m}^{Coul}(L_i) -   
             \sum_{L_i = 0}^{L_{i}^{max}}{\hat \beta}_{\ell m}^{Coul} (L_i),
\end{eqnarray}
where the first term on the right hand side, is the pure Coulomb 
breakup amplitude which for
the case where one of the outgoing fragments
is uncharged, can be expressed analytically in terms of the 
bremsstrahlung integral (see Ref. \cite{cha00}). Therefore, only
two terms, with reasonable upper limits, are required to be evaluated
by the partial wave expansion in Eq.~(\ref{eq19_5}). 
\section{Calculations on $^{11}$Be}

\subsection{Structure model and optical potentials}
The wave function, $u_\ell(r)$, appearing in the structure term, $Z_\ell$ [Eq.~(\ref{eqst})],
has been calculated by adopting a single particle potential
model in the same way as in Ref. \cite{cha00,fuku}. 
The ground state of $^{11}$Be was considered to be a predominantly $s$-state
with a $2s_{1/2}$ valence neutron coupled to the $0^+$ $^{10}$Be core
[$^{10}$Be $\otimes$ $2s_{1/2}\nu$] with a one-neutron separation energy 
of 504 keV and a spectroscopic factor of 0.74 \cite{aum}.
The single particle wave function was constructed by assuming the 
valence neutron-$^{10}$Be interaction to be of Woods-Saxon type
whose depth was adjusted to reproduce the corresponding value
of the binding energy with fixed values of the radius and diffuseness
parameters (taken to be 1.236 fm and 0.62 fm, respectively). 
This gave a potential depth of 59.08 MeV, a root mean square (rms) radius
for the valence neutron of 7.07 fm, and a rms radius for $^{11}$Be
of 2.98 fm when the size of the $^{10}$Be core was taken to be 2.28 fm.
 The neutron-target optical potentials used by us 
were extracted from the global set of Bechhetti-Greenlees
(see, e.g,~\cite{per76}), while those used for the $^{10}$Be-target
(\cite{fuku,cha03}) system are shown in Table I. Following~\cite{typ01},
we have used the sum of these two potentials for the
$^{11}$Be-target channel. We found that values of $L_i^{max}$ of 500 and 400
for Pb and Ti targets, respectively, provided very good convergence of the
corresponding partial wave expansion series [Eq.~(\ref{eq17_5})]. The 
local momentum wave vectors are evaluated at a distance, $R$ = 10 fm
in all the cases,
and their directions are taken to be same as that of asymptotic momenta.
More details on the validity
of this approximation can be found in the appendix of Ref. \cite{cha03}.

\begin{table}[h]
\caption{Optical potential parameters for the $^{10}$Be-target
interaction. Radii are calculated with the $r_jt^{1/3}$ convention.}
\begin{center}
\begin{tabular}{|c|c|c|c|c|c|c|}
\hline
system &  $V_r$ & $r_r$& $a_r$&  $W_i$& $r_i$& $a_i$\\
       & (MeV) & (fm) & (fm) & (MeV) & (fm) & (fm)      \\       
\hline
$^{10}$Be--$^{44}$Ti &70 & 2.5 & 0.5 & 10.0 & 1.5 & 0.50 \\
$^{10}$Be--$^{208}$Pb & 50 & 1.45 & 0.8 & 57.9 & 1.45 & 0.8  \\
\hline
\end{tabular}
\end{center}
\end{table}

\subsection{Neutron energy distribution}
\begin{figure}
\centering
\includegraphics[width=.75\textwidth]{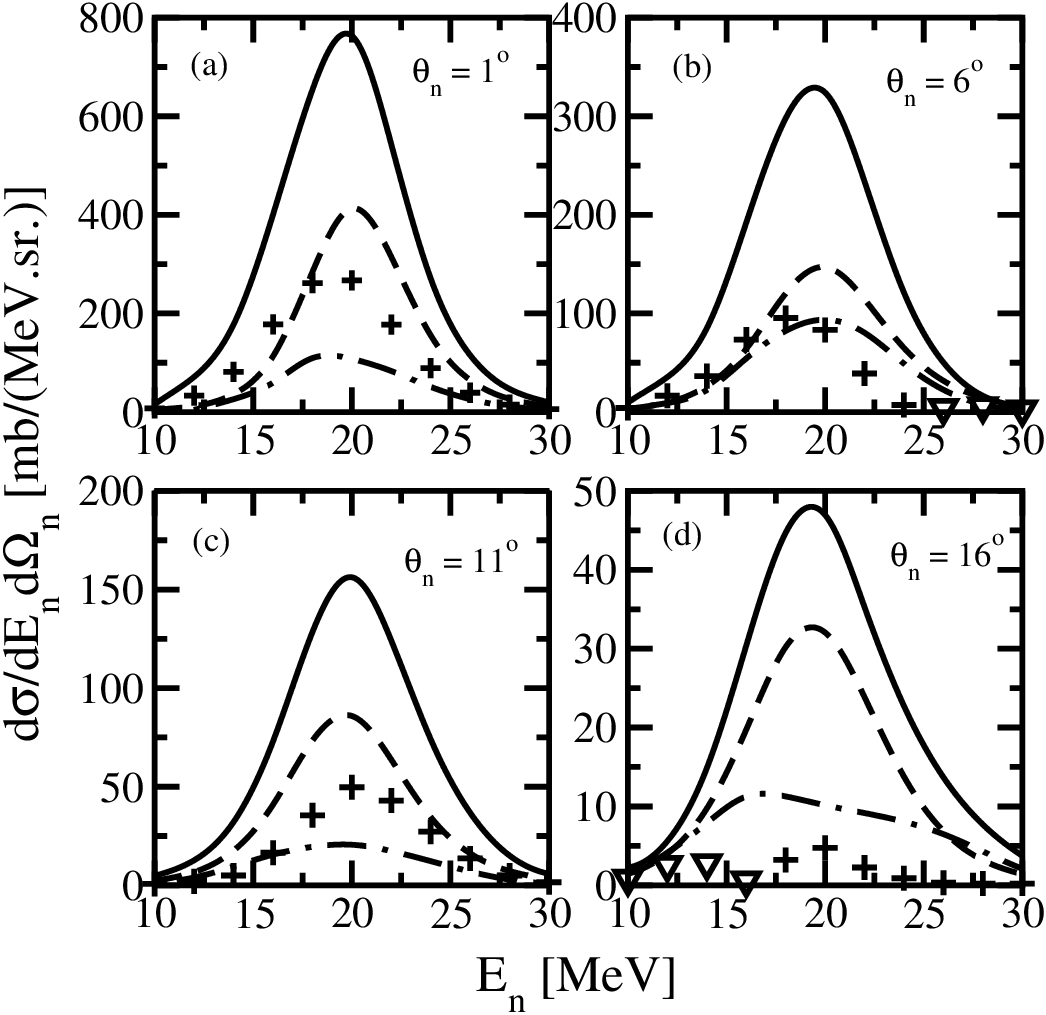}
\caption{\label{fig1} 
Neutron energy distribution for the breakup reaction $^{11}$Be on $^{44}$Ti 
at the beam energy of 20 MeV/nucleon, for $\theta_n =$
$1^{\circ}$, $6^{\circ}$, $11^{\circ}$ and $16^{\circ}$. 
The dashed and dot-dashed
lines represent the pure Coulomb and nuclear contributions, respectively,
while total contributions are shown by solid 
lines. The plus signs
and the inverted triangles represent the magnitudes of the 
positive and negative interference terms, respectively.} 
\end{figure}

It had been observed earlier that the CNI terms were dependent more on exclusive observables
than on inclusive ones mainly due to the fact that multiply integrated quantities (theoretically) washed away the
effect of interferences. Calculations of the double differential cross section (neutron energy distribution)
for two forward neutron emission angles in the breakup of $^{11}$Be on Au 
at the beam energy of 41 MeV/nucleon, showed that the CNI terms were dependent on angles and energies 
of the outgoing neutron \cite{cha03}.
Their magnitudes were nearly equal to those of the nuclear breakup contributions
which led to a difference in the incoherent and coherent sums of the
Coulomb and nuclear contributions underlying thus the importance of
those terms. 
In this sub-section we shall present results of our calculations for the neutron energy distribution for the breakup 
of $^{11}$Be on a medium mass target at various beam energies and neutron emission angles ($\theta_n$). In all these cases the
core emission angle ($\theta_{^{10}{\rm Be}}$) is integrated from $0^{\circ}$ to $30^{\circ}$.

In Fig.\ 1, we plot the neutron energy distribution for the breakup of $^{11}$Be on $^{44}$Ti 
at the low beam energy of 20 MeV/nucleon, for
$\theta_n =$ $1^{\circ}$, $6^{\circ}$, $11^{\circ}$ and $16^{\circ}$. 
The dashed and dot-dashed
lines represent the pure Coulomb and nuclear contributions, respectively,
while total contributions are shown by solid 
lines. The plus signs
and the inverted triangles represent the magnitudes of the 
positive and negative interference terms, respectively. At all neutron emission angles the Coulomb 
breakup terms are more than the nuclear ones at this low beam energy. We also see that the interferences at this beam energy are 
constructive, in general, and that at the neutron angles of
$1^{\circ}$ and $11^{\circ}$, in Figs. 1(a) and 1(c), respectively the magnitude of the interference terms are more 
than the nuclear terms for nearly all energies of the outgoing neutron.

\begin{figure}
\centering
\includegraphics[width=.75\textwidth]{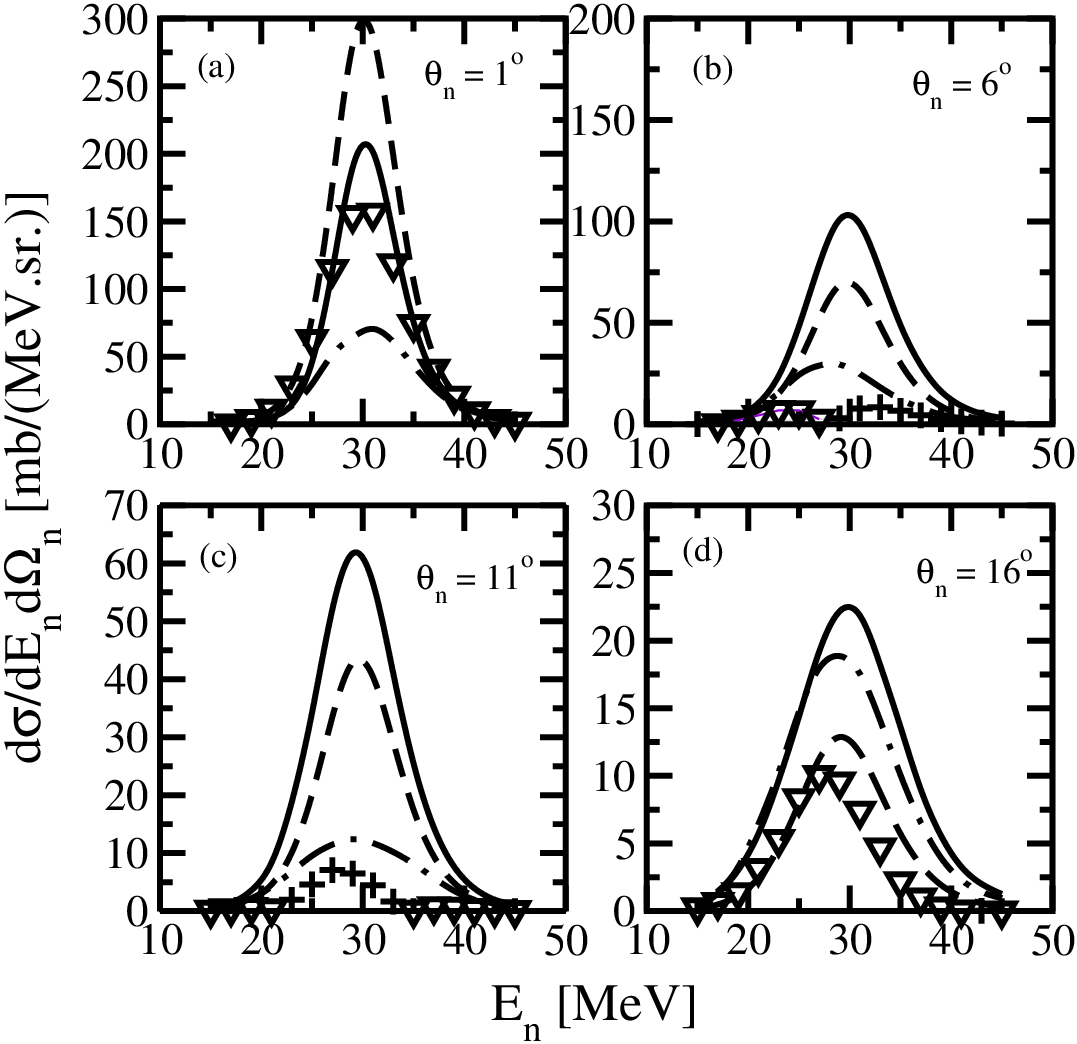}
\caption{\label{fig2} 
Neutron energy distribution for the breakup reaction $^{11}$Be on $^{44}$Ti 
at the beam energy of 30 MeV/nucleon, for $\theta_n =$
$1^{\circ}$, $6^{\circ}$, $11^{\circ}$ and $16^{\circ}$. 
The dashed and dot-dashed
lines represent the pure Coulomb and nuclear contributions, respectively,
while total contributions are shown by solid 
lines. The plus signs
and the inverted triangles represent the magnitudes of the 
positive and negative interference terms, respectively.} 
\end{figure}

A similar calculation performed at a higher beam energy is shown 
in Fig.\ 2, where we plot the neutron energy distribution for the breakup of $^{11}$Be on $^{44}$Ti 
at the beam energy of 30 MeV/nucleon, for $\theta_n =$
$1^{\circ}$, $6^{\circ}$, $11^{\circ}$ and $16^{\circ}$. 
Pure Coulomb and nuclear contributions are shown by dashed and dot-dashed
lines, respectively,
while total contributions are shown by solid 
lines. 
The plus signs
and the inverted triangles represent the magnitudes of the 
positive and negative interference terms, respectively. At 30 MeV/nucleon, incident beam energy, we
see that the Coulomb terms are larger than the nuclear ones at smaller neutron emission angles [Figs. 2(a-c)],
while at larger angles the nuclear part begins to dominate. The importance of the interference terms is highlighted in
Fig. 2(a), where we see that the destructive interference terms not only cancel out the nuclear
terms, but also reduces the Coulomb terms so that the coherent total sum is less than the Coulomb terms. At 
$\theta_n = 16^{\circ}$ [Fig. 2(d)] we see that the destructive CNI terms nearly cancel out the Coulomb terms, especially
from neutron energies of 15 MeV to 27 MeV, and the nuclear terms are sole contributers at these energies.

\begin{figure}
\centering
\includegraphics[width=.75\textwidth]{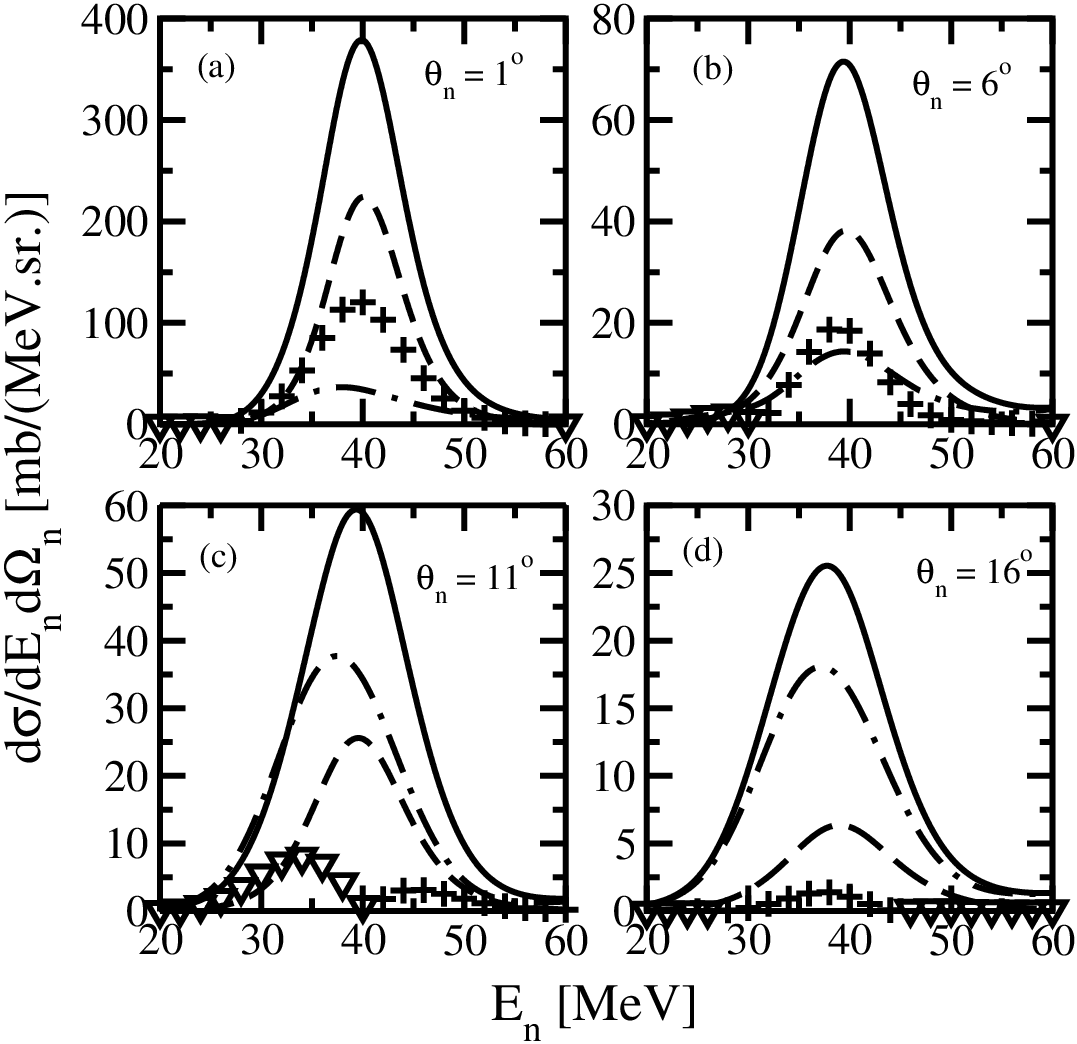}
\caption{\label{fig3} 
Neutron energy distribution for the breakup reaction $^{11}$Be on $^{44}$Ti 
at the beam energy of 40 MeV/nucleon, for $\theta_n =$
$1^{\circ}$, $6^{\circ}$, $11^{\circ}$ and $16^{\circ}$. 
The dashed and dot-dashed
lines represent the pure Coulomb and nuclear contributions, respectively,
while total contributions are shown by solid 
lines. The plus signs
and the inverted triangles represent the magnitudes of the 
positive and negative interference terms, respectively.} 
\end{figure}

In Fig.\ 3, we plot the neutron energy distribution for the breakup of $^{11}$Be on $^{44}$Ti 
at the beam energy of 40 MeV/nucleon, for $\theta_n =$
$1^{\circ}$, $6^{\circ}$, $11^{\circ}$ and $16^{\circ}$. 
The dashed and dot-dashed
lines represent the pure Coulomb and nuclear contributions, respectively,
while total contributions are shown by solid 
lines. The plus signs
and the inverted triangles represent the magnitudes of the 
positive and negative interference terms, respectively. 
We
see that the Coulomb terms are larger than the nuclear ones at smaller neutron emission angles [Figs. 3(a,b)],
while at larger angles the nuclear part begins to dominate. The interference is generally constructive at smaller
neutron angles, often being larger or almost equal to the individual nuclear terms ($\theta_n = 1^{\circ}, 6^{\circ}$), while
at $\theta_n = 11^{\circ}$, especially from neutron energies of 20 MeV to 35 MeV, the destructive CNI terms nearly cancel out the Coulomb terms,  and the nuclear terms are sole contributers to the total cross section.

Our results, thus, indicate that the CNI terms are not only dependent on energies and angles of the outgoing fragments, 
they are also dependent on the incident beam energy. It would indeed be quite interesting if more exclusive cross section measurements
could be made at low beam energies, where the effect of the CNI terms were found to be substantial, in a future experiment.
\subsection{Relative energy spectra}

\begin{figure}
\centering
\includegraphics[width=.75\textwidth]{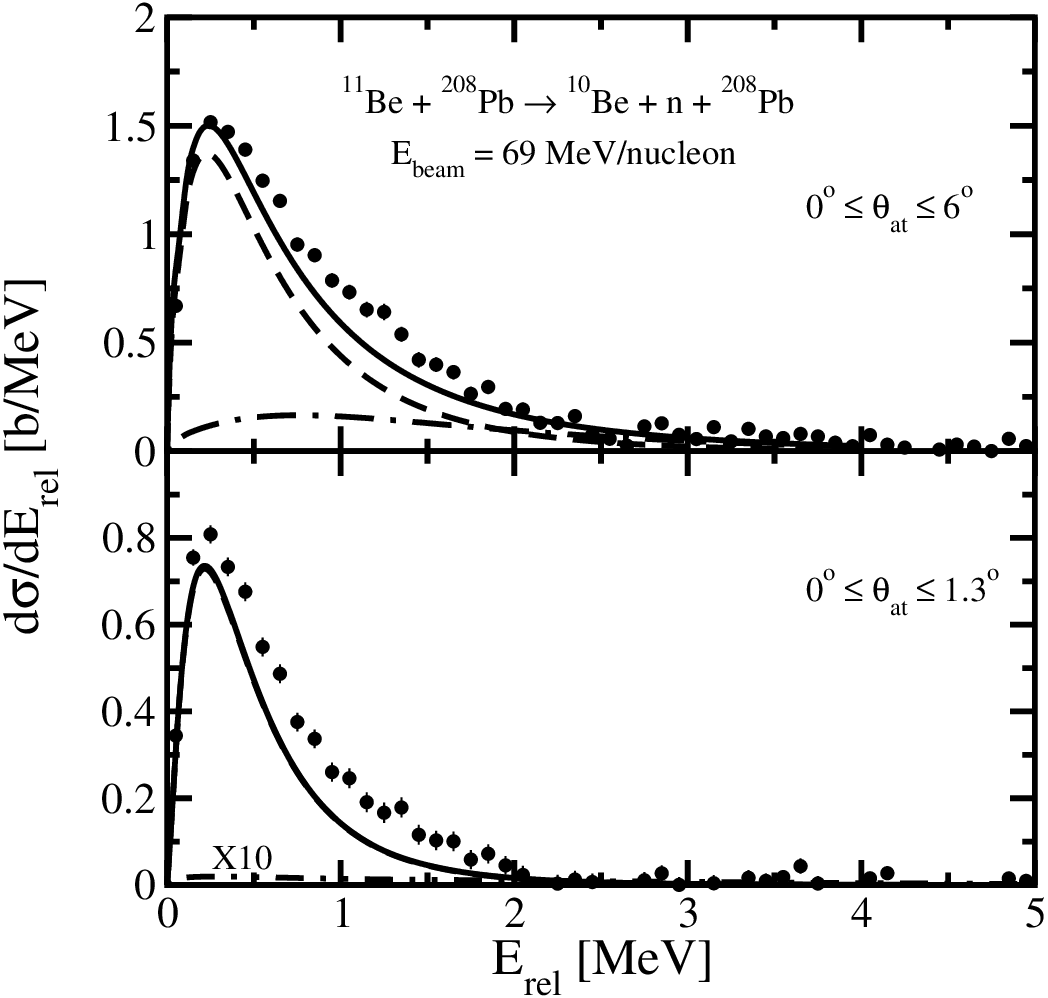}
\caption{\label{fig4}Relative energy spectra for the breakup of $^{11}$Be 
on a $^{208}$Pb target at 69 MeV/nucleon, incident beam energy, for different angular
ranges of $\theta_{at}$. The dashed and dot-dashed
lines represent the pure Coulomb and nuclear breakup 
contributions, respectively, while total contributions are
shown by solid lines. In the upper panel, integration over 
$\theta_{at}$ has been done in the range of 
$0^{\circ}$--$6^{\circ}$, while in the lower panel $\theta_{at}$
is integrated in the range of $0^{\circ}$--$1.3^{\circ}$. Experimental data are from 
Ref. \cite{fuku}. For more details see the text.
}
\end{figure}

The relative energy spectrum of the fragments (neutron and $^{10}$Be) 
emitted in the breakup of $^{11}$Be on   
$^{208}$Pb target at the beam energy of 69 MeV/nucleon
is shown in Fig. 4, for different angular ranges of the projectile c.m scattering
angle ($\theta_{at}$). The relative
angle between the fragments ($\theta_{n-{^{10}{\rm Be}}}$) 
has been integrated from $0^{\circ}$ to $180^{\circ}$. The dashed and dot-dashed
lines represent the pure Coulomb and nuclear contributions, respectively,
while total coherent contributions are shown by solid 
lines. The experimental data are from Ref. \cite{fuku}.

In the upper panel of Fig. 4, $\theta_{at}$--integration has been done in the range of 
$0^{\circ}$--$6^{\circ}$. The pure
Coulomb contributions dominate the cross sections around the peak value,
while at larger relative energies
the nuclear breakup is important.
This was also observed in Refs.~\cite{typ01,cha03} and was 
 attributed to the different energy dependence of the two
contributions. In Ref.~\cite{cha03}, however, the $\theta_{at}$--integration was
done from $0^{\circ}$--$40^{\circ}$ mainly to account for all nuclear contributions coming from small impact parameters.

The Coulomb breakup contribution has a
long range and it shows a strong energy dependence. 
The number of virtual photons increases for small excitation energies
and hence the cross sections
rise sharply at low excitation energies. After a certain value of this
energy the cross sections decrease due to setting in of the
adiabatic cut-off. In contrast, the nuclear breakup occurs when the projectile and the 
target nuclei are close to each other. Its magnitude, which is determined
mostly by the geometrical conditions, has a weak dependence
on the relative energy of the outgoing fragments beyond a certain
minimum value.  
The coherent sum of the Coulomb and nuclear contributions provides a
good overall description of the experimental data. 

The lower panel of Fig. 4, shows the relative energy spectra in which 
the $\theta_{at}$--integration was
done from $0^{\circ}$--$1.3^{\circ}$. This is well below the grazing angle (= $3.8^{\circ}$)
for the reaction and consequently 
the Coulomb contribution in this case dominates 
over the nuclear part (which in fact is multiplied by a factor of 10 to make
it visible in the figure). Thus, the dashed and solid lines in the lower panel of Fig. 4, almost coincide with each other.
This will also be reflected in the total one-neutron removal cross section which we present in the next sub section.

\subsection{Total one-neutron removal cross section}
\begin{table}[ht]
\caption{Total one-neutron removal cross section, various contributions
from pure Coulomb and pure nuclear breakups, and their incoherent
sum for $^{11}$Be breakup 
on $^{208}$Pb , at beam energy of 69 MeV/nucleon for two different angular ranges of $\theta_{at}$.}
\begin{center}
\begin{tabular}{|c|c|c|c|c|c|c|c|}
\hline
$\theta_{at}$ &Total & Pure Coulomb & Pure nuclear & Incoherent sum& 
    Expt.~\cite{fuku}\\
  (deg.) &(b)    & (b)          & (b)          & (b)           &
    (b)               \\
\hline
 $0^{\circ}$--$6^{\circ}$& 1.534 & 1.191 & 0.367 & 1.558 & 1.790$\pm$0.110(syst)$\pm$0.020(stat)\\
  $0^{\circ}$--$1.3^{\circ}$& 0.489 & 0.484 & 0.005 & 0.489 & --  \\
\hline
\end{tabular}
\end{center}
\end{table}

In Table II, we show the
contributions of pure Coulomb and pure nuclear breakup mechanisms to
the total one-neutron removal cross sections in the breakup of
$^{11}$Be on $^{208}$Pb for two different angular ranges of $\theta_{at}$, at the beam energy of 69 MeV/nucleon.
The incoherent sum shown in the penultimate column of the table is obtained by simply adding the pure
Coulomb and pure nuclear cross sections.

For the breakup of $^{11}$Be on Pb in the $\theta_{at}$--range of $0^{\circ}$--$6^{\circ}$,
Coulomb breakup accounts for most ($\approx$ 78{\%}) of the total cross section, while in the
angular range of $0^{\circ}$--$1.3^{\circ}$ (well below the grazing angle) it accounts for almost all of 
the total cross section.

The total one-neutron removal cross section on Pb 
does not seem to be affected by the CNI terms. This is because the CNI terms 
manifest themselves explicitly in more exclusive measurements,
like double differential cross sections than in
quantities like total cross sections.

\section{Summary and Conclusions}

In this paper, we have investigated the beam energy dependence of the Coulomb-nuclear interference terms in the breakup of 
$^{11}$Be on a medium mass $^{44}$Ti target  
and have also calculated the relative energy spectra for the breakup of $^{11}$Be on a heavy $^{208}$Pb target at 69 MeV/nucleon
for two different angular ranges of the projectile center of the mass scattering angle. The calculations were performed within 
the fully quantum mechanical framework of post-form DWBA,
where pure Coulomb, pure nuclear as well as their
interference terms were treated consistently within the same framework.
In this theory, both the Coulomb and nuclear interactions between 
the projectile and the target nucleus were treated to all orders, but 
the fragment-fragment interaction was treated in the first order.
The full ground state wave function of the projectile corresponding to
any orbital angular momentum structure enters as an input to this theory.

The exact post-form DWBA breakup amplitude was simplified with the LMA
and the validity of the approximation was verified by calculating several reaction observables
in the breakup of $^{11}$Be in the Coulomb and nuclear fields of several targets, in different mass ranges, in Refs. \cite{cha02,cha03}. 
Recently, there have been attempts to calculate the exact post-form DWBA without the LMA, for pure Coulomb breakup, with momentum space 
Coulomb wave functions \cite{zadro}. While, this is indeed an welcome step, its practical applicability to calculate a wide variety
of reaction observables is still an open question, particularly because it is numerically very intensive and time consuming. Thus the LMA 
still has practical applications, in this respect, particularly because along with its ability to factorise the Coulomb breakup
amplitude it is also able to treat Coulomb and nuclear breakup on a single footing. Nevertheless, efforts are in progress \cite{rc_new} for the
calculation of the post-form DWBA breakup amplitude without the LMA in a more analytic and less numerically intensive way than
using momentum space Coulomb wave functions.

We calculated the neutron energy distributions for the breakup 
of $^{11}$Be on $^{44}$Ti at various beam energies and neutron emission angles. At 20 MeV/nucleon, beam energy, the Coulomb breakup
accounted for more of the cross section than nuclear breakup and the CNI terms were constructive, in general. The importance of the CNI
terms were again highlighted by the calculation at the beam energy of 30 MeV/nucleon, where the CNI terms at low neutron emission angles,
not only cancel out the nuclear
terms, but also reduces the Coulomb terms so that the coherent total sum is less than the Coulomb terms. At 40 MeV/nucleon, beam energy,
interference was generally constructive at smaller
neutron angles, often being larger or almost equal to the individual nuclear terms ($\theta_n = 1^{\circ}, 6^{\circ}$), while
at $\theta_n = 11^{\circ}$, especially from neutron energies of 20 MeV to 35 MeV, the destructive CNI terms nearly cancels out the Coulomb terms,  
and the nuclear terms are sole contributers to the total cross section.
Our results, thus, indicate that the CNI terms are not only dependent on energies and angles of the outgoing fragments,
 they are also dependent on the incident beam energy. It would indeed be quite interesting if in a future experiment exclusive cross section measurements
could be made at low beam energies where the effect of the CNI terms were found to be substantial.

Calculations were also performed for the relative energy spectrum of the fragments (neutron and $^{10}$Be) 
emitted in the breakup of $^{11}$Be on $^{208}$Pb target at the beam energy of 69 MeV/nucleon, for different 
angular ranges of the projectile c.m scattering
angle -- $0^{\circ}$ to $6^{\circ}$ and $0^{\circ}$ to $1.3^{\circ}$. In the former angular range pure
Coulomb contributions dominate the cross sections around the peak value,
while at larger relative energies
the nuclear breakup is important. In the latter range, which is well below the grazing angle, for the reaction
Coulomb breakup dominates over the nuclear part. 

The total one-neutron removal cross section was found not to be affected by the CNI terms as they manifest themselves explicitly in more exclusive measurements,
like double differential cross sections than in
quantities like total cross sections.

The full quantal theory of one-neutron halo
breakup reactions, applied in this paper, can  also be used to
describe the $(a,b\gamma)$ reaction provided the inelastic breakup
mode is also calculated within this theory, which is expected to be straightforward. 
Furthermore, the theoretical method outlined in this paper rely on the nonrelativistic 
Schr\"odinger equation which in our opinion should be viewed only as an adequate
starting point. There have been some attempts
to use effective field methods to study halo 
nuclei \cite{bert02}. This is indeed a very new field and would be quite interesting to pursue
in view of experiments of halo breakup at very high beam energies for which data have been taken
at GSI, Darmstadt.

\acknowledgments
It is a pleasure to thank Prof. R. Shyam and Prof. A. Vitturi for many interesting discussions.
Thanks, also to Prof. T. Nakamura
for providing the experimental data shown in the Fig. 4
in a tabular format.
 
\end{document}